\documentclass[11pt,a4paper]{article}
\usepackage[T1]{fontenc}
\usepackage[utf8]{inputenc}
\usepackage{lmodern}
\usepackage{placeins}
\usepackage[margin=1in]{geometry}
\usepackage{microtype}
\usepackage{amsmath}
\usepackage{amssymb}
\usepackage{graphicx}
\usepackage{float}
\usepackage{booktabs}
\usepackage{tabularx}
\usepackage{caption}
\usepackage[numbers,square,sort&compress]{natbib}

\usepackage[hidelinks]{hyperref}

\title{\vspace{-3em}\LARGE\textbf{AgenticMobility}\\[-0.1em]
An Open-Source Platform for Multi-AI-Agent Mobility Dynamics\\[0.3em]}

\author{%
Jiaheng Xiao\textsuperscript{1}\quad 
Jiangtao Han\textsuperscript{2}\quad 
Jiachen Wang\textsuperscript{3}\quad
Honghai Li\textsuperscript{4}\quad\\
Lijun Sun\textsuperscript{5}\quad 
Kay Axhausen\textsuperscript{6}\quad 
Chang Liu\textsuperscript{7}\quad 
Zhengbing He\textsuperscript{1\thanks{Corresponding author: \href{mailto:he.zb@hotmail.com}{he.zb@hotmail.com}}}
\\[0.75em]
\small\textsuperscript{1}Faculty of Science and Engineering, University of Nottingham Ningbo China, China\\
\small\textsuperscript{2}College of Management and Economics, Tianjin University, China\\
\small\textsuperscript{3}Beijing Key Laboratory of Traffic Engineering, Beijing University of Technology, China\\
\small\textsuperscript{4}Research Institute of Highway, Ministry of Transport, China\\
\small\textsuperscript{5}Department of Civil Engineering, McGill University, Canada\\
\small\textsuperscript{6}Institute for Transport Planning and Systems, ETH Zurich, Switzerland\\
\small\textsuperscript{7}School of Economics, University of Nottingham Ningbo China, China\\
}
\date{}

\renewcommand{\abstractname}{Abstract}
\renewenvironment{abstract}
  {\par\begin{center}\bfseries\abstractname\end{center}\normalsize}
  {\par}
\begin{document}

\maketitle
\vspace{-1.5em}

% Abstract
% \input{sections/abstract}
\begin{abstract}
Understanding mobility dynamics and the effects of urban transportation management policies requires examining how individual travel decisions, traffic conditions, and policy interventions influence one another over time.
We present AgenticMobility, an open-source experimental platform for studying how mobility AI agents and other AI agents interact, adapt through repeated experience, and collectively shape mobility outcomes.
Persistent agents retain identities and relevant states across rounds, with explicit interfaces defining information access, admissible actions, and decision schedules; linked records trace decisions, execution, and feedback.

In the current implementation, large language model (LLM) agents act as travelers choosing routes and departure times and as a manager periodically adjusting link charges within MATSim.
A preliminary demonstration spans 100 simulated days in each of five configurations, with one run per configuration and a shared simulation seed.
In the joint traveler--manager run, all 1,179,300 finalized target decisions satisfied action constraints; execution checks verified implementation of accepted choices and preservation of personal memories across days.
Trajectories show stable route choices alongside departure-time adjustments and differences between controlled-trip and population-wide outcomes.

The platform supports experiments on individual adaptation, collective outcomes, and management interventions by varying information access, retained experience, and management feedback.
Source code is available at \url{https://github.com/Agentic-City-Research/Agentic-Mobility}.
\end{abstract}

\vspace{1.5em}

\newpage

\tableofcontents

\newpage

\section{Introduction}
\label{sec:introduction}
Mobility systems evolve through repeated decisions by travelers, traffic managers, and service operators.
Travelers choose when and how to travel under individual activity requirements, while sharing limited network capacity.
Their choices jointly generate congestion and travel experiences that influence later decisions \cite{qi2023investigating,Qi2024RouteChoices,liu2020day}.
Managers respond to aggregate conditions by changing rules or costs, which in turn alter the conditions faced by travelers \cite{rambha2016dynamic, cantarella1995dynamic,ye2015learning}.
Understanding this feedback requires examining how individual adaptation and management interventions jointly affect participating agents and the wider population.

Multi-agent simulation provides an important experimental approach to studying travel behavior, collective traffic dynamics, and transportation management \cite{horni2016matsim}.
Conventional approaches typically represent agent decisions through predefined rules or mathematical models \cite{mahmassani1987boundedly, lou2010robust}.
These representations offer explicit behavioral assumptions, but specifying rules or functional relationships for diverse motivations and adaptive responses becomes increasingly difficult as behavioral heterogeneity and situational complexity grow.

Rapid advances in generative AI, particularly large language models (LLMs), offer a new opportunity to model agents capable of human-like reasoning and decision-making.
LLM-based agents can draw on role-specific attributes, available information, and past experiences to generate responses to changing situations \cite{park2023generative,piao2025agentsociety}.
Beyond individual decisions, interactions among LLM agents can also produce collective outcomes, including population-level conventions and biases \cite{ashery2025emergent}.

Recent studies have begun to incorporate LLM agents into transportation modeling, from individual travel decisions to urban mobility simulation \cite{li2024more, liu2025llm, yan2024opencity}.
\cite{wang2025agentic} introduced LLMTraveler and studied day-to-day route choices through comparisons with laboratory observations and experiments in multi-origin--destination congestion settings.
\cite{liu2026gatsim} developed GATSim to simulate urban mobility by connecting generative residents with adaptive travel processes.
Related work provides conceptual frameworks for transportation agents and implementations linking LLM-generated plans to mobility execution \cite{liu2024transport,patwary2026bridging}.
These studies demonstrate the potential of LLM agents for modeling travel behavior, but their evaluations focus on traveler decisions or specific tasks, often with small populations or limited LLM participation.
Real transportation systems involve large, heterogeneous populations with different roles, so plausible individual choices alone do not establish how their interactions shape system dynamics.

We introduce \emph{AgenticMobility}, an open-source experimental platform for studying \emph{multi-AI-agent mobility dynamics}: how mobility AI agents and other AI agents adapt through repeated interaction and shape collective mobility outcomes.
Persistent identities and states connect decisions across rounds, while linked records support inspection of individual and collective trajectories.
The framework allows roles and decision types to be specified for each experiment; the current MATSim-based implementation comprises route and departure-time choices for travelers and periodic link-charge adjustments for a manager.

The paper makes three contributions: (i) a framework for studying collective mobility dynamics through repeated interactions among mobility AI agents and other AI agents with different transportation roles; (ii) an open-source experimental platform coupling LLM agents with MATSim, with linked records that make decisions, execution, and feedback traceable across rounds; and (iii) operational validation and illustrative experiments showing how the platform supports investigation of individual adaptation, management feedback, and differences between controlled-trip and population-wide outcomes.
Source code is available at \url{https://github.com/Agentic-City-Research/Agentic-Mobility}.

The remainder of the paper is organized as follows.
Section~\ref{sec:system_framework} presents the interaction framework, and Section~\ref{sec:reproducibility} describes its implementation and reproducibility.
Section~\ref{sec:validation} reports execution validation and illustrative experiments.
Section~\ref{sec:scope} discusses research opportunities and current limitations, and Section~\ref{sec:conclusion} concludes the paper.

\section{System Framework}
\label{sec:system_framework}

\begin{figure}[htbp]
    \centering
    \includegraphics[width=\linewidth, trim=0cm 4cm 0cm 1cm, clip]{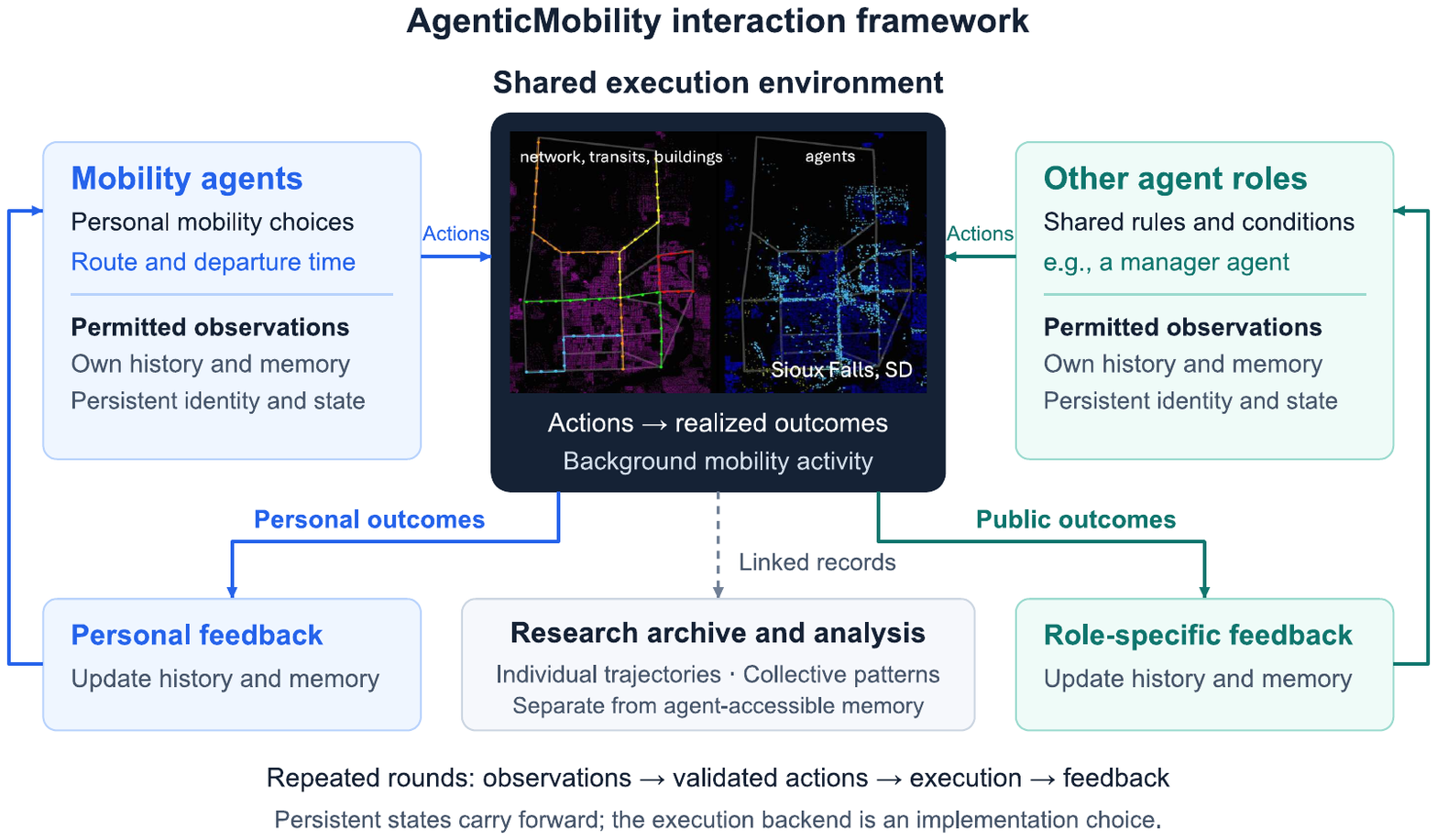}
    \caption{AgenticMobility interaction framework.
Role-specific actions and feedback connect agents through a shared execution environment.
Persistent states link successive rounds; a separate archive records the interaction process.
The traveler and manager roles illustrate one study setting rather than a fixed set of platform roles.}
    \label{fig:framework}
\end{figure}

Mobility AI agents form the main simulated population in AgenticMobility and interact with other AI agents undertaking complementary roles such as traffic management or service operation.
Each study specifies their roles, information access, action spaces, and decision schedules.
Role-specific observations inform decisions, accepted actions are executed in a shared transportation environment, and feedback updates persistent states for subsequent decisions.
Figure~\ref{fig:framework} summarizes this process.
A round comprises a coordinated cycle of decisions, execution, feedback, and state updates.

The framework separates three responsibilities: agent decision-making, environmental execution, and research observation.
This separation allows researchers to specify what an agent can know and do, preserve the backend's operational constraints, and inspect the relationship between a submitted action and its consequences.
The interfaces below define these responsibilities independently of the daily schedule and MATSim backend used in the current implementation.

\subsection{Roles, Observations, and Persistent State}
\label{sec:roles-state}

Each agent has a persistent identity, role attributes, an admissible action space, and an evolving state.
Its role determines which decisions it can make and which information it receives.
Persistence links successive decisions to the same agent identity and retained state \cite{park2023generative}.

Observations combine static descriptions, available announcements, and earlier outcomes according to role-specific access rules.
Realized outcomes become available only after execution; private records, shared announcements, and public summaries follow the study's visibility rules.

State representations and update rules are role-specific, with observed outcomes distinguished from generated interpretations.
A separate research archive retains interaction records without giving agents unrestricted access.

\subsection{Action and Execution Interfaces}
\label{sec:actions}

Agents submit structured actions containing an identifier, a decision time, and role-specific fields.
The action interface specifies admissible values, units, scope, and effective timing.
Validation checks these constraints before an action is accepted.
Generated explanations are retained separately from executable fields.

An execution adapter applies accepted actions through operations supported by the backend.
The environment combines these actions with background activity and shared resource constraints, then returns outcome records.
Accepted actions, their application, and realized outcomes remain distinct.
A valid action does not guarantee the intended outcome: congestion, resource competition, and earlier activities can alter its realized consequences.

The adapter preserves action scope and timing and records application, completion, and feedback; the backend determines environmental dynamics.

Table~\ref{tab:interface-contracts} summarizes the information that must be specified at each interface to connect agent decisions, environmental execution, and state updates.

\begin{table}[H]
    \centering
    \small
    \caption{Information requirements for the platform
interfaces.
Entries specify the meaning and scope of exchanged information; implementation-specific fields may vary.}
    \label{tab:interface-contracts}
    \linespread{1.2}\selectfont
\begin{tabularx}{\linewidth}{
        @{}
        >{\raggedright\arraybackslash}p{0.2\linewidth}
        >{\raggedright\arraybackslash}X
        @{}
    }
        \toprule
        Interface & Required information \\
        \midrule
        Observation
        & Receiving agent and role, decision round,
          information sources, and access permissions. \\

       Action
& Acting agent, submitted values, affected entities
  or resources, effective timing, and validity constraints. \\

Outcome
& Relevant actions and execution round, application
  and completion status, and realized results. \\

        State update
        & Previous persistent state, newly available
          feedback, and updated state. \\
        \bottomrule
    \end{tabularx}
\end{table}

\subsection{Interaction Timing, Feedback, and Traceability}

The coordination protocol specifies when each role acts, when its actions take effect, and when the resulting information becomes available to others.
Roles may act at different frequencies, and their actions can change the conditions faced by others.
Decisions use only information available before they are made; outcomes enter subsequent decision contexts after execution.

\begin{figure}[H]
    \centering 
    \includegraphics[width=\linewidth, trim=0cm 0.3cm 0cm 0.2cm, clip]
{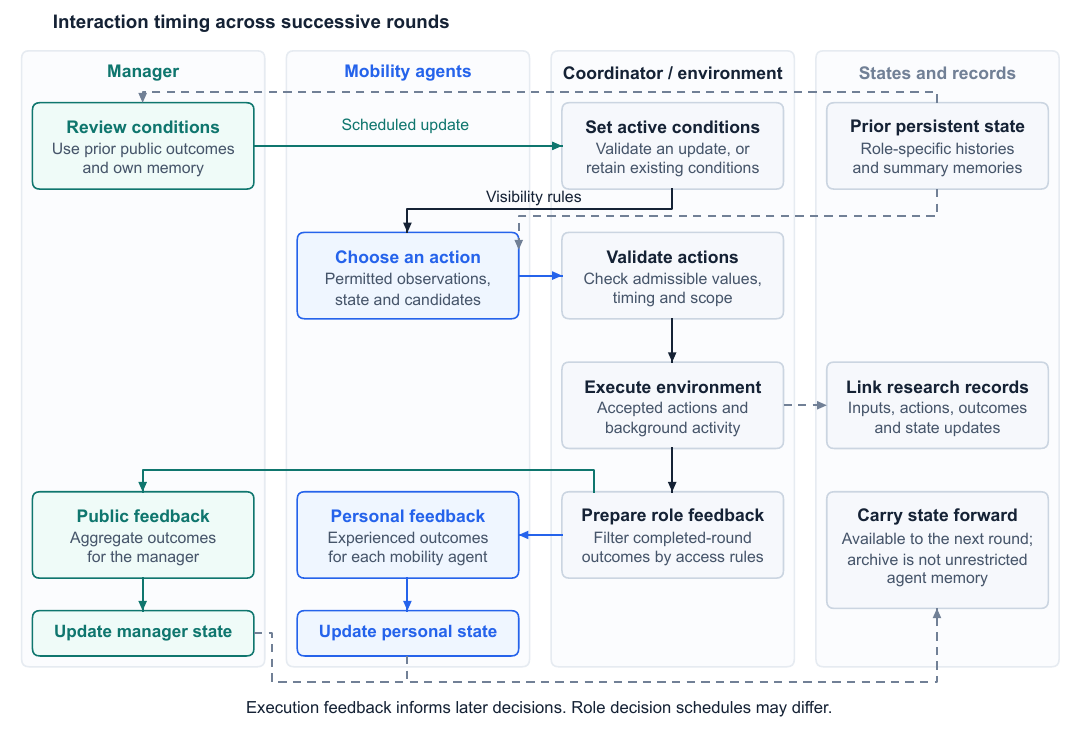}
\caption{Illustrative interaction timing and state
updates across successive rounds for traveler and
manager roles. Solid arrows indicate action and
feedback flows; dashed arrows indicate state transfer
and research-record linkage. Role schedules are
study-specific, and the research archive remains
separate from agent-accessible memory.}
    \label{fig:interaction_timing}
\end{figure}

After execution, permitted outcomes and summaries update agent states for subsequent decisions.
Figure~\ref{fig:interaction_timing} summarizes the sequence.

For each decision, the research archive links the supplied observations, generated output, validation result, applied action, realized outcome, and subsequent state update.
Records also identify relevant shared conditions, including any active policy, and any execution or update failures.
Appendices~\ref{app:concrete-protocol} and~\ref{app:implementation} specify the implemented daily schedule and failure-handling rules.

\section{Platform Implementation}
\label{sec:reproducibility}

\subsection{Implemented Mobility Setting}
\label{sec:implemented-setting}

The current implementation uses MATSim \cite{horni2016matsim} as the execution backend, with each round corresponding to one simulated day.
In this setting, mobility AI agents act as travelers choosing routes and departure times for fixed target trips, while a traffic manager takes on a complementary role by periodically adjusting link charges.
Both roles use LLMs for decision-making and are referred to below as travelers and the manager.
Trips outside the LLM-controlled target set retain MATSim behavioral mechanisms, preserving full-day activity chains and their effects on the shared environment.

The manager is instructed to reduce full-day delay measured from completed link traversals.
Unlike population-wide car time, this objective excludes waiting before network entry; both measures are reported separately.

\subsection{Agent Decision-Making and Memory}
\label{sec:agent-implementation}

Table~\ref{tab:agent_config} summarizes how
the traveler and manager roles instantiate
the interaction framework introduced in
Section~\ref{sec:system_framework},
with distinct observations, actions,
persistent states, and decision schedules.
These configurations are specific to the
present demonstration rather than fixed
platform restrictions.

\begin{table}[H]
\centering
\small
\caption{Role-specific agent mechanisms and settings in AgenticMobility.}
\label{tab:agent_config}
\setlength{\tabcolsep}{4pt}
\linespread{1.2}\selectfont
\begin{tabularx}{\linewidth}{
    @{}>{\raggedright\arraybackslash}p{0.13\linewidth}
    >{\raggedright\arraybackslash}X
    >{\raggedright\arraybackslash}X
    >{\raggedright\arraybackslash}X
    >{\raggedright\arraybackslash}X@{}}
\toprule
& \multicolumn{2}{c}{Traveler} & \multicolumn{2}{c}{Manager} \\
\cmidrule(lr){2-3}\cmidrule(l){4-5}
Component & Mechanism & Setting & Mechanism & Setting \\
\midrule
Role
& Target-trip decision maker & 11,793 agents
& Road pricing manager & 1 agent \\

Action
& Route choice; departure-time adjustment
& $\leq 4$ routes; $-90$ to $+30$ min, 15-min steps
& Link-charge adjustment & $\leq 40$ car links \\

Observation
& Personal trip outcomes and permitted toll information & ---
& Public network conditions and aggregate outcomes & --- \\

Memory
& Factual history; personal-experience summary
& 3-day history; 500-character summary
& Policy history; public-feedback summary
& Summary $\leq 1{,}000$ words \\

Decision schedule
& Trip decisions & Daily
& Policy reviews & Days~20, 50, and 80 \\

Memory update
& Personal-memory update & Daily
& Public-feedback memory update & Daily \\
\bottomrule
\end{tabularx}
\end{table}

Traveler decisions additionally depend on synthetic
preference profiles and schedule references.
Informed travelers receive current toll
announcements, whereas uninformed travelers
do not, but retain their own payment experiences.
The manager has access to public network information
and aggregate outcomes, but not to private traveler
profiles or memories.

Traveler choice, personal-memory updates, manager policy decisions, and manager-memory updates are configured as separate LLM tasks.
Table~\ref{tab:llm-settings} reports the generation settings recorded for the combined configuration.
Model identifiers do not fix remote service versions or weights.
Thinking is disabled; effective temperature and top-$p$ values are unspecified in the recorded settings.

\begin{table}[H]
\centering\small
\caption{Recorded model settings for LLM Traveler + Manager. Token limits
are per-response maxima.}
\label{tab:llm-settings}
\linespread{1.2}\selectfont
\begin{tabularx}{\linewidth}{@{}>{\raggedright\arraybackslash}Xllr@{}}
\toprule
Task & Model identifier & Output format & Max. tokens \\
\midrule
Traveler choice & \texttt{qwen3.5-9b} & JSON schema & 1,024 \\
Personal memory & \texttt{qwen3.5-9b} & JSON schema & 1,024 \\
Manager policy & \texttt{deepseek-flash} & JSON object & 8,192 \\
Manager memory & \texttt{deepseek-flash} & JSON object & 2,048 \\
\bottomrule
\end{tabularx}
\end{table}

\paragraph{Experience retention and memory updates.}
\label{sec:memory-updates}

Completed traveler outcomes enter a rolling factual
history, while a separate LLM task updates the
summary memory using the previous summary and new
feedback. Outcomes leaving the factual window remain
available in subsequent decision contexts only if
retained in the summary.

\begin{figure}[!tbp]
\centering
\includegraphics[width=\linewidth,height=0.44\textheight,keepaspectratio]
{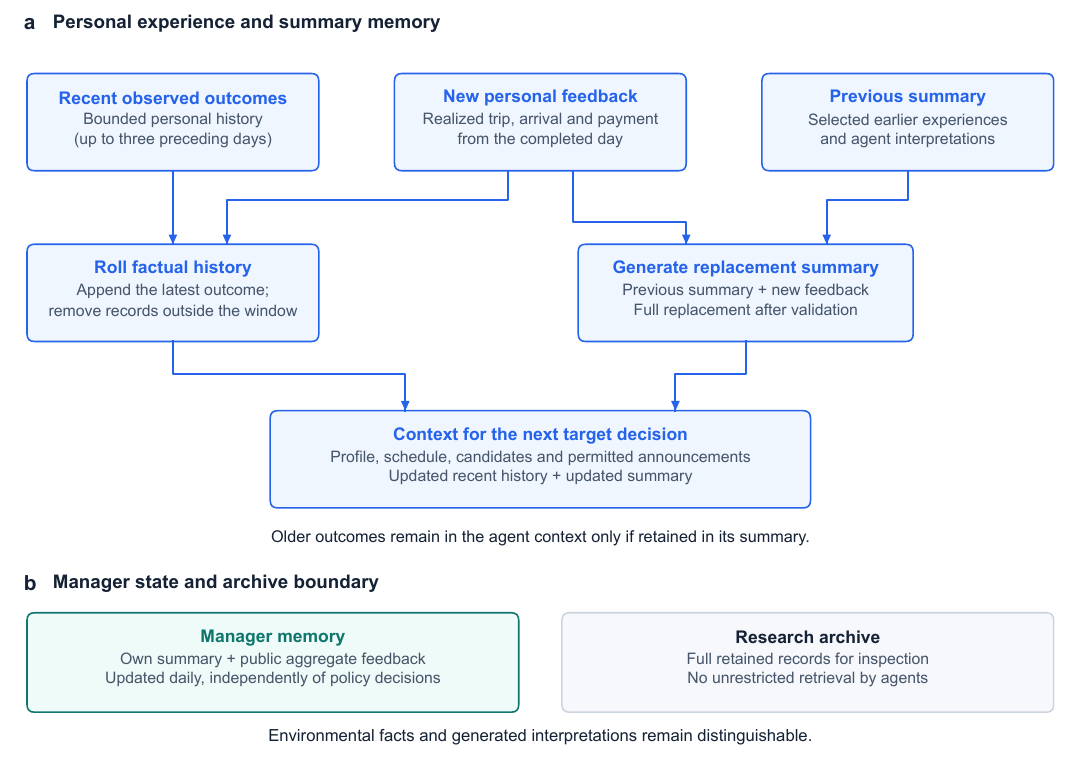}
\caption{Experience retention and state updates. New outcomes update the
rolling factual history and supply summary generation.
Validated summaries enter later inputs.
Manager memory updates daily, independently of policy decisions, and the research archive remains outside unrestricted agent access.}
\label{fig:memory_update}
\end{figure}

Manager memory is updated daily using public feedback
and policy history, independently of scheduled policy
reviews. Observed outcomes remain distinct from
generated summaries, as shown in Figure~\ref{fig:memory_update}.
Appendix~\ref{app:decision-models} provides the MNL specification and prompt entry points.

\subsection{Implementation and Execution Workflow}
\label{sec:execution-workflow}

\begin{figure}[!tbp]
\centering
\includegraphics[
    width=\linewidth,
    height=0.46\textheight,
    keepaspectratio
]{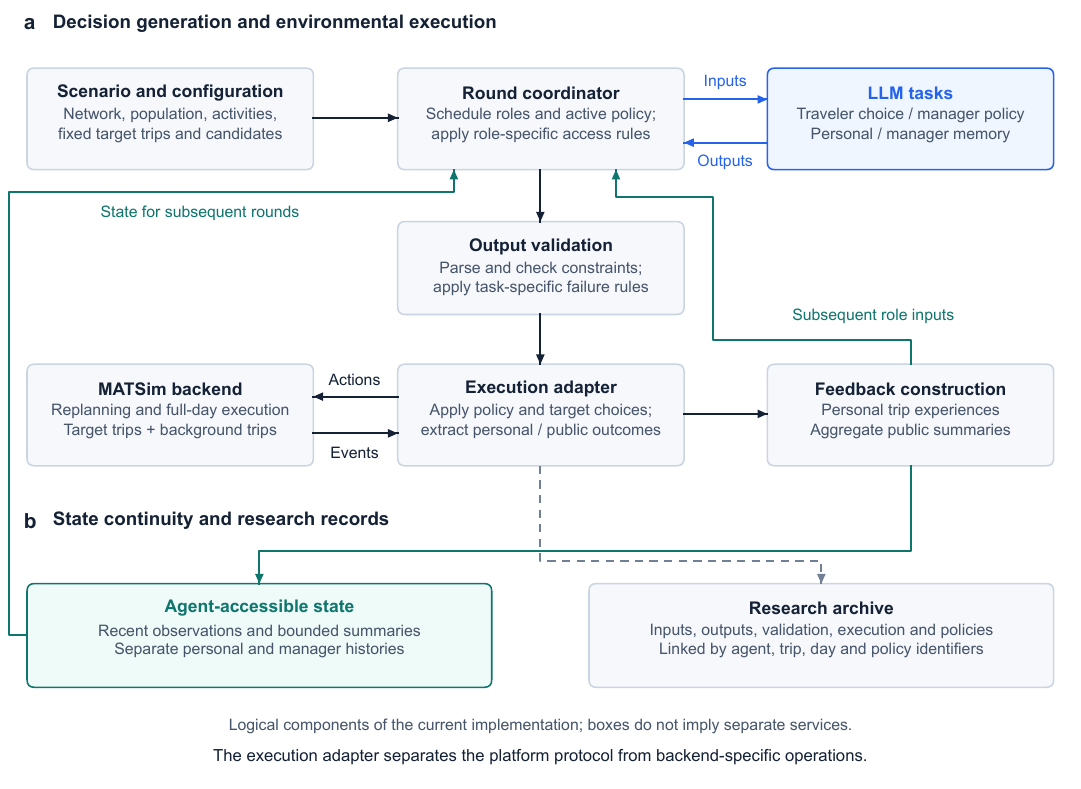}
\caption{
Components and data flows in the implementation.
The coordinator manages agent decisions, validation,
and state updates, while the execution adapter
connects these mechanisms to the MATSim backend.
Returned events support role-specific feedback,
with agent states and research records maintained
separately.
}
\label{fig:implementation_dataflow}
\end{figure}

The implementation separates general coordination
mechanisms from backend-specific execution procedures.
The coordinator constructs role-specific contexts,
schedules decision and memory tasks, validates
generated actions against configured constraints,
and manages persistent states and research records.
An execution adapter connects these mechanisms
to the simulation backend, translating accepted
actions into executable operations and extracting
feedback for subsequent decisions.
Figure~\ref{fig:implementation_dataflow}
illustrates this workflow; its components represent
logical responsibilities rather than separate
deployed services.

In the current MATSim implementation, the execution
adapter applies the active charging policy and
writes accepted traveler choices into target plans
after background replanning and before execution.
MATSim then simulates the full day, and its event
records are used to construct personal outcomes
and public summaries according to role-specific
access rules.

Selected and realized departure times are
distinguished because earlier activities and trips
may delay execution.
Appendix~\ref{app:implementation} specifies
detailed input construction, validation rules,
and failure handling.

\subsection{Open-Source Release and Reproduction Scope}
\label{sec:release-scope}

The AgenticMobility implementation is available at \href{https://github.com/Agentic-City-Research/Agentic-Mobility}{AgenticMobility}.

The documented installation check runs a small population for two simulated days without LLM calls.
Enabling LLM-controlled roles additionally requires a compatible model service.
Table~\ref{tab:release-materials} summarizes the released materials and their intended uses.

\begin{table}[H]
    \centering
    \small
    \caption{Materials provided with the AgenticMobility
release and their intended uses.}
    \label{tab:release-materials}

    \linespread{1.2}\selectfont
\begin{tabularx}{\linewidth}{
        @{}
        >{\raggedright\arraybackslash}p{0.35\linewidth}
        >{\raggedright\arraybackslash}X
        @{}
    }
        \toprule
        Material & Intended use \\
        \midrule

        Source code, scenario inputs, and configuration guides
        & Build and run the platform and configure new
          experiments. \\

        Aggregate outputs and run summaries
        & Inspect daily trajectories and recalculate
          reported summary measures. \\

        Execution-audit records and selected case excerpts
        & Inspect recorded action checks and examples linking
          decision inputs, actions, and outcomes. \\

        Run metadata and artifact hashes
        & Identify the recorded runs and trace available
          executable and input artifacts. \\

        \bottomrule
    \end{tabularx}
\end{table}

Regenerating historical trajectories requires the corresponding executable or source revision, inputs, prompts, configuration, and model-service settings.
A simulation seed alone does not determine remote model responses.
Appendix~\ref{app:archive-scope} describes the run-specific artifacts and reproduction scope.

\section{Demonstration and Evaluation}
\label{sec:validation}

The demonstration examines execution consistency and illustrates research uses through action audits, interaction traces, and collective trajectories.

\subsection{Demonstration Setup}
\label{sec:demonstration-setup}

The demonstration uses an enriched Sioux Falls scenario \cite{chakirov2014sioux} with 16,822 simulated people and 11,793 fixed target trips. The simulated population represents a fixed 20\% sample of the 84,110 people in the input scenario \cite{ben-dorPopulationDownscalingMultiagent2021}. This sampling rate was selected to make repeated LLM-agent decisions computationally feasible while retaining a population large enough to demonstrate interactions within the shared transportation network. To account for population downsampling, the MATSim flow and storage capacity factors were set to 0.2 and $0.2^{0.75}$, respectively \cite{mersiniScalingMATSimTraffic2026}. 

In the LLM traveler configurations, 11,793 mobility AI agents each control one target trip; the remaining 5,029 people and all non-target trips retain MATSim behavioral mechanisms.
Full-day activity chains and background mobility remain active in all configurations.
The configurations exercise the platform with neither, either, or both LLM roles enabled, alongside a multinomial logit (MNL) reference mechanism for target-trip choices (Table~\ref{tab:conditions}). MNL and LLM travelers use the same precomputed BFS-LE route candidate file \cite{rieser2013route}.
All use MATSim for environmental execution.

\begin{table}[H]
\centering\small
\caption{Demonstration configurations. Trips outside the controlled target set
retain MATSim behavioral mechanisms.}
\label{tab:conditions}
\linespread{1.2}\selectfont
\begin{tabularx}{\linewidth}{@{}>{\raggedright\arraybackslash}p{0.29\linewidth}
>{\raggedright\arraybackslash}X >{\raggedright\arraybackslash}p{0.17\linewidth}@{}}
\toprule
Configuration & Target-trip decisions & LLM manager \\
\midrule
MATSim & MATSim plan selection and replanning & Disabled \\
MNL & Constrained joint route--departure choice & Disabled \\
LLM Traveler & LLM route--departure choice & Disabled \\
MATSim + Manager & MATSim plan selection and replanning & Enabled \\
LLM Traveler + Manager & LLM route--departure choice & Enabled \\
\bottomrule
\end{tabularx}
\end{table}

Each configuration completed 100 simulated days using simulation seed 268.
MNL and LLM travelers select from the same fixed route and departure-time candidates, whereas MATSim agents retain their own plan-selection and replanning mechanisms.
The MNL coefficients are uncalibrated.
There is one run per configuration, and successive days share persistent states.
Table~\ref{tab:configuration} reports the principal settings; Appendix~\ref{app:demonstration} and Appendix~\ref{app:configuration} provide supplementary conditions and scenario details.

Days are indexed from 0 to 99, following the logs.
Manager review opportunities occur on Days~20, 50, and 80.
Positive charging begins on Day~20 in MATSim + Manager and on Day~50 in LLM Traveler + Manager.
In the figures, \emph{LLM} abbreviates LLM Traveler, and \emph{LLM + Manager} denotes the combined configuration.

The controlled action space permits up to four candidate routes per origin--destination pair and departure offsets from $-90$ to $+30$ minutes relative to the original plan, in 15-minute increments.
Options are clipped against activity and simulation boundaries; 2,780 targets have only one candidate route.
The original departure remains available, and arrival deadlines are not hard feasibility constraints.

\begin{table}[!tbp]
\centering\small
\caption{Principal settings for the demonstration. Strategy entries are
configured weights, not probabilities inferred from their sum.}
\label{tab:configuration}
\linespread{1.2}\selectfont
\begin{tabularx}{\linewidth}{@{}>{\raggedright\arraybackslash}p{0.34\linewidth}
>{\raggedright\arraybackslash}X@{}}
\toprule
Setting & Value \\
\midrule
MATSim version & 2026.0 \\
Population / simulation seeds & 6021 / 268 \\
Initial strategy weights & \texttt{ChangeExpBeta}: 0.7;
\texttt{ReRoute}: 0.1; \texttt{TimeAllocationMutator}: 0.1 \\
Innovation cutoff & Log Day~90; existing-plan selection continues \\
Charging interval & $[0,108000)$ seconds \\
Manager policy reviews & Log Days~20, 50, and 80 \\
Manager action limits & At most 40 distinct car links; finite nonnegative
charge per entry; no preset price cap or increment \\
Recent personal observations & Up to three preceding days \\
Personal summary limit & 500 characters \\
Manager summary limit & 1,000 English words; target length 400--600 words,
with no enforced minimum \\
Information groups & 5,897 informed and 5,896 uninformed travelers \\
\bottomrule
\end{tabularx}
\end{table}

\FloatBarrier
\subsection{Evaluation Measures}
\label{sec:measurement}

\paragraph{Individual choice changes.}
Let $I$ be the fixed cohort of $N=11,793$ target trips, with $\rho_{i,d}$ denoting the complete ordered link sequence of the selected route and $\tau_{i,d}$ the selected departure time on day $d$.
Route and departure-time change rates are
\begin{equation}
 R_d=\frac{1}{N}\sum_{i\in I}\mathbf{1}(\rho_{i,d}\ne\rho_{i,d-1}),
 \qquad
 A_d=\frac{1}{N}\sum_{i\in I}\mathbf{1}(\tau_{i,d}\ne\tau_{i,d-1}).
 \label{eq:behavioral-adjustments}
\end{equation}
The rates cover 99 adjacent-day transitions.
Multi-route rates restrict both numerator and denominator to the 9,013 targets with multiple fixed candidates.
MATSim agents are not restricted to those candidate sets.
Equations define proportions, which are multiplied by 100 when reported as percentages.

\paragraph{Aggregate route-flow turnover.}
If $n_{o,r,d}$ is the number of targets in OD pair $o$ using complete route $r$ on day $d$, net turnover is
\begin{equation}
 Q_d=\frac{\sum_{o,r}|n_{o,r,d}-n_{o,r,d-1}|}{2N}.
\end{equation}
The same percentage conversion applies.
Individual switches can offset one another in route counts, so $Q_d$ and $R_d$ measure different forms of change.

\paragraph{Time and schedule outcomes.}
Let $H_d^{\mathrm{car}}$ be total daily car time in person-hours.
Mean daily car time per simulated person is
\begin{equation}
 \overline T_d^{\mathrm{car}}=\frac{60H_d^{\mathrm{car}}}{16,822}.
\end{equation}
The unit is minutes per person.
Total car time covers all car legs and includes waiting before network entry.
Mean target-trip time covers the 11,793 controlled trips, while work late-arrival rates use the fixed denominator of 11,332 work targets.
All planned car trips completed in the reported runs.

Selected departure offsets are relative to the original plans; negative values indicate earlier departures.
Early-arrival measures do not assign a monetary value to early arrival or schedule inconvenience.
Completed-link delay, used in manager feedback, excludes waiting before network entry and is not interchangeable with total car time.
None of these time measures alone represents a calibrated welfare measure that combines travel time, payments, and schedule inconvenience.

\paragraph{Aggregation windows.}
Outcome summaries average daily values over all 100 days.
Choice-change rates average 99 transitions.
Relative differences are calculated from unrounded window means; figure smoothing uses trailing seven-day means.
Consecutive days and people sharing a run are not independent run-level replications.
The Day~90 innovation cutoff is a configured change in the background mechanism, not an inferred convergence time.

\subsection{Execution and Feedback Consistency}
\label{sec:validation-execution}

All five configurations completed 11,793 target trips and 33,644 car legs per day, with no stuck agents.
Work-arrival measures cover the same 11,332 targets throughout.
Table~\ref{tab:execution-audit} distinguishes full-record checks of the combined configuration from sampled inspection of decision inputs.

\begin{table}[htbp]
\centering\small
\caption{Execution audit scope and results. Action checks concern finalized
outputs after validation and repair.
The information-access result applies to the inspected samples.}
\label{tab:execution-audit}
\linespread{1.2}\selectfont
\begin{tabularx}{\linewidth}{@{}>{\raggedright\arraybackslash}p{0.23\linewidth}
>{\raggedright\arraybackslash}p{0.29\linewidth}
>{\raggedright\arraybackslash}X@{}}
\toprule
Check & Coverage & Recorded result \\
\midrule
Action constraints and execution
& 1,179,300 finalized target decisions in the combined run
& Valid route and departure choices; accepted actions linked to applied plans
and completed routes \\
Personal state continuity
& 100-day personal-memory chains in the combined run
& Continuous chains; finalized summaries within the configured limit \\
Manager decisions and state
& 3 policy decisions and 100 daily memory updates in the combined run
& Valid policies; all memory updates completed \\
Information access
& 4,000 contexts: 2,000 from each LLM traveler configuration
& No current announcements disclosed to uninformed travelers and no
network-wide traffic information disclosed \\
\bottomrule
\end{tabularx}
\end{table}

Output repair affected 0.206\% of finalized traveler-decision records and 0.250\% of personal-memory records in the combined run.
One manager-memory response required JSON repair.
Appendix~\ref{app:audit-cost} gives exact repair counts and the task-level call breakdown.

These checks establish execution and state continuity; memory accuracy and explanation faithfulness require separate evaluation.

\paragraph{Computational workload.}
Table~\ref{tab:cost} reports the inference workload for all five configurations.
These totals describe inference workload rather than monetary cost or end-to-end runtime.

\begin{table}[H]
\centering\small
\caption{LLM inference workload over complete 100-day runs. Calls include
retries and repairs; tokens are service-reported totals.}
\label{tab:cost}
\linespread{1.2}\selectfont
\begin{tabularx}{\linewidth}{@{}Xrr@{}}
\toprule
Configuration & API calls & Tokens \\
\midrule
MATSim & 0 & 0 \\
MNL & 0 & 0 \\
LLM Traveler & 2,362,964 & 5,713,538,990 \\
MATSim + Manager & 109 & 872,774 \\
LLM Traveler + Manager & 2,364,090 & 5,451,720,152 \\
\bottomrule
\end{tabularx}
\end{table}

\subsection{Recorded Interaction Traces}
\label{sec:interaction-trace}

\paragraph{Traveler choice and subsequent experience.}
Figure~\ref{fig:execution-trace} follows synthetic traveler \texttt{10010\_2} in the combined configuration.
On Day~60, the agent selected route \texttt{c1} and the earliest available departure after receiving recent outcomes that included a late arrival on Day~59.
The applied action and completed route matched the choice, and the outcome entered the rolling personal history for subsequent decisions.

\begin{figure}[htbp]
\centering\small
\fbox{\begin{minipage}{0.93\linewidth}
\textbf{Observation before the Day~60 decision.} Four candidate routes, nine departure options, a 09:00 arrival reference, personal outcomes from Days~57--59, and current toll announcements.
Network-wide conditions are unavailable.
The Day~59 trip arrived at 09:10:48.
\par\smallskip\centerline{$\downarrow$}\smallskip
\textbf{Structured action.} Route \texttt{c1}, with selected departure at 05:56:43, the earliest option and 90 minutes before the original plan.
\par\smallskip\centerline{$\downarrow$}\smallskip
\textbf{Execution and outcome.} The realized departure matches the selected time, and the completed route matches \texttt{c1}.
Car time is 21.97 minutes, destination arrival is 06:33:36, and payment is zero.
Door-to-door time is 36.88 minutes and uses different boundaries from car time.
\par\smallskip\centerline{$\downarrow$}\smallskip
\textbf{Subsequent decision.} On Day~61, the traveler selects \texttt{c0} at 05:56:43.
The Day~60 outcome is available in the rolling personal history.
\end{minipage}}
\caption{A recorded traveler interaction linking permitted observations,
an accepted action, realized execution, and a subsequent decision.
Route, departure time, and shared conditions vary together.}
\label{fig:execution-trace}
\end{figure}

Because route, departure time, and shared conditions changed together, it does not identify their separate effects.
A second case distinguishes execution from explanation: traveler \texttt{10042\_2} selected and executed a departure at 07:46:19 on Day~61, while its generated reason stated 07:49:39.
The cases distinguish execution fidelity from explanation accuracy; they do not estimate error prevalence.

\paragraph{Manager policy, feedback, and memory.}
In the combined run, the Day~50 manager input contained public observations through Day~49, the active zero-charge policy, and the summary written after Day~49.
The accepted decision introduced charges of 1.5 scenario units per entry on links \texttt{43\_3} and \texttt{43\_4}, effective on Day~50.
After execution, public feedback entered the daily memory task, whose replacement summary was carried into the Day~51 memory-update input.
The memory update did not itself change the policy.

At the next scheduled policy decision on Day~80, the manager retained the same charges.
A new policy version was recorded, but the price function remained unchanged.
Appendix~\ref{app:manager-trace} documents the linked inputs, policy application, feedback, and state updates.

\subsection{Illustrative Collective Dynamics}
\label{sec:illustrative-use}

\paragraph{Individual and population outcomes.}
In these runs, LLM Traveler has the lowest mean daily car time per person, 19.90\% below MATSim, whereas LLM Traveler + Manager has the shortest mean target-trip time.
Relative to LLM Traveler, the combined configuration records 6.63\% lower target-trip time and 14.62\% higher daily car time per person.
Differences already exist before positive charging, so this contrast cannot be attributed to the manager alone.

\begin{figure}[!htbp]
    \centering
    \includegraphics[width=\linewidth, keepaspectratio ]{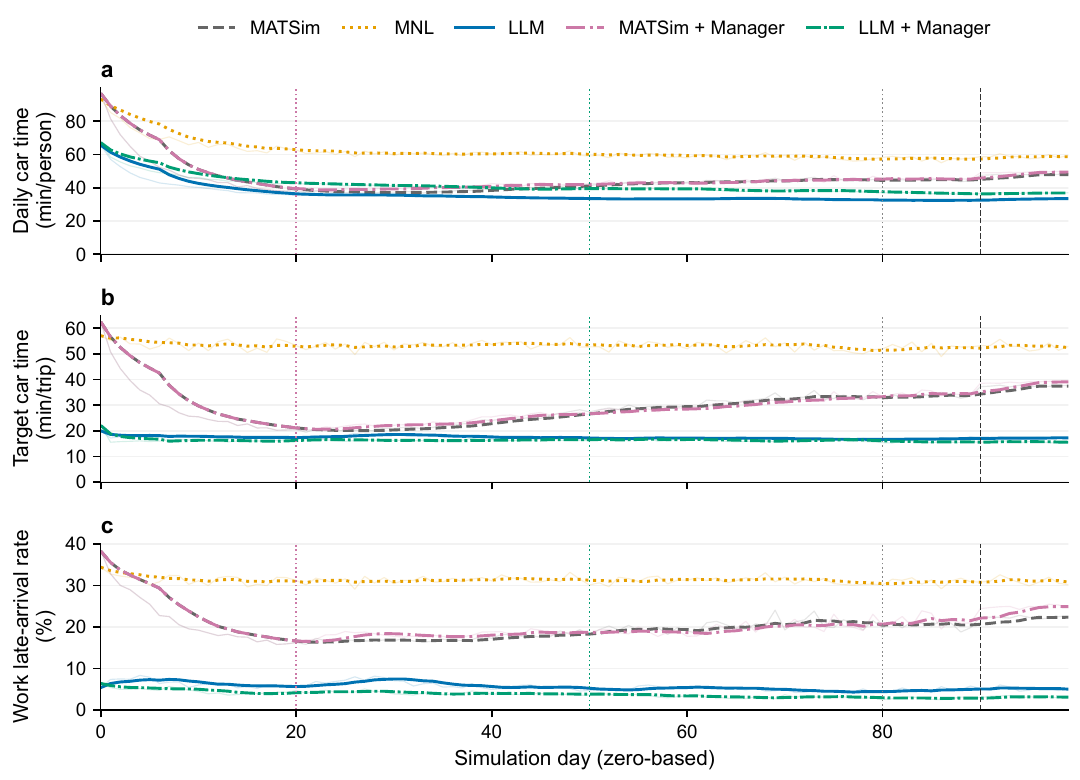}
    \caption{Individual and population mobility outcomes across five
configurations: (a) mean daily car time per person, (b) mean target-trip car time, and (c) work-target late-arrival rate.
Thin lines show daily values; thick lines show trailing seven-day means.
Positive charging begins on Day~20 for MATSim + Manager and Day~50 for LLM + Manager.
Day~90 marks the MATSim innovation cutoff.
Each configuration has one seed-268 run.}
    \label{fig:outcomes}
\end{figure}

Figure~\ref{fig:outcomes} shows the corresponding daily trajectories.
A reduction in controlled-trip time need not coincide with an improvement in the wider population's daily travel, making the outcome boundary part of the research question.

\paragraph{Individual adjustment and aggregate change.}
Route choices change less frequently than departure times in the two LLM traveler configurations.
Across 99 adjacent-day transitions, overall route-switching rates are 3.08\% for LLM Traveler and 2.19\% for the combined configuration, while departure-time change rates are 29.45\% and 28.04\%.
These rates include 2,780 targets with only one candidate route.
Restricting the route-switching calculation to the 9,013 targets with multiple routes gives 4.03\% and 2.86\%, respectively.

\begin{figure}[!htbp]
    \centering
    \includegraphics[width=\linewidth, keepaspectratio ]{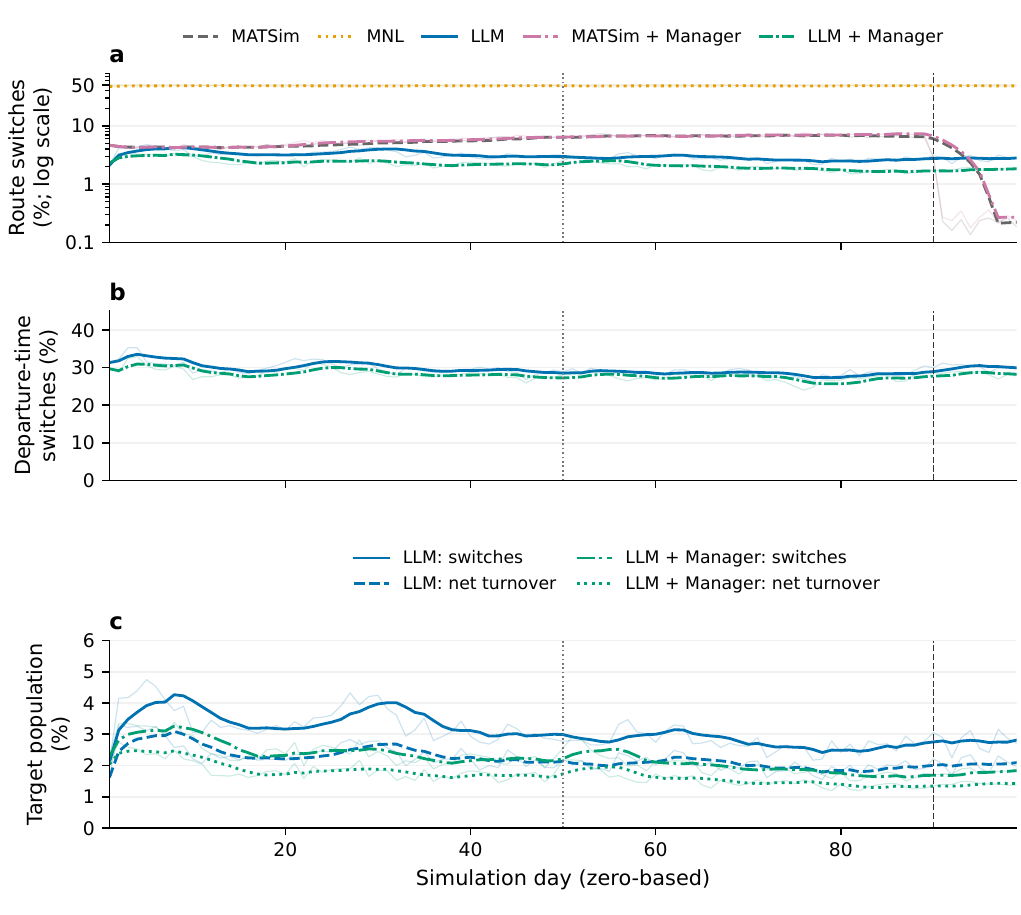}
    \caption{Individual choice changes and aggregate route-flow turnover.
(a) Route-switching rates across five configurations on a logarithmic scale.
(b) Departure-time changes in the two LLM traveler configurations.
(c) Individual route switching and net turnover of OD-specific route counts.
Rates use all 11,793 targets.
Thin lines show daily values; thick lines show trailing seven-day means.
Day~50 marks charging in LLM + Manager, and Day~90 marks the MATSim innovation cutoff.}
    \label{fig:behavior}
\end{figure}

Figure~\ref{fig:behavior} compares individual switching with aggregate route-flow turnover.
Individual changes can offset one another within an origin--destination pair, leaving aggregate route counts relatively stable.
The decline in MATSim route switching near Day~90 also reflects the configured cutoff for route and activity-time innovation; existing-plan selection continues after the cutoff.

LLM Traveler and the combined configuration select departures an average of 34.96 and 39.65 minutes earlier than the original plans.
These schedule adjustments must be considered alongside travel time and punctuality when assessing overall travel burden.
The route-choice stability and departure adjustments also raise a behavioral question: under what information and preference conditions do agents respond primarily by changing when they travel rather than which route they use?
Appendix~\ref{app:supplementary} reports the numerical summaries.

\FloatBarrier

\section{Discussion}
\label{sec:scope}

The demonstration motivates further research on agent behavior, collective dynamics, and management feedback.
Both mobility AI agents and other AI agents are subjects of investigation.

\subsection{Individual Adaptation and Collective Mobility Dynamics}
The observed route-choice stability and departure-time adjustments motivate studies of how information and retained experience shape adaptation \cite{mahmassani1986experiments}.
Experiments can vary announcement access, factual-history windows, and summary-memory mechanisms while keeping candidate actions and other conditions comparable \cite{liu2017interactive}.
Such studies can investigate how apparently beneficial individual responses accumulate into outcomes for the controlled cohort and the wider population.

\subsection{Traveler-Manager Feedback}
Experiments can examine how policy review frequency, announcement timing, and retained experience affect management responses and subsequent traffic conditions.
Controlled interventions, repeated runs, and clearly specified outcome boundaries are needed to examine when management feedback improves travel conditions and how its benefits and burdens are distributed.

\subsection{AI-Agent Capabilities and State Accuracy}
Repeated decisions across transportation roles provide a setting for studying AI-agent consistency and adaptation under spatial, temporal, and information constraints.
The recorded explanation inconsistency illustrates why action validity, execution fidelity, and the accuracy of generated interpretations must be evaluated separately.
Likewise, a continuous memory chain confirms that state is passed between rounds, but does not establish that the summary retains relevant facts or supports appropriate decisions \cite{maharana2024evaluating, wu2024longmemeval}.
Comparisons across models and memory mechanisms can hold roles, information, actions, and execution conditions fixed while assessing consistency, factual accuracy, adaptation, and collective outcomes.
Objective performance and behavioral realism require separate evaluation.

\subsection{Behavioral Calibration and Network-Level Validation}
Calibration can examine how assumed preference profiles, population composition, memory mechanisms, and other model settings jointly shape individual behavior and network outcomes.
Observed choices and behavioral distributions can constrain individual-level representations, while traffic counts, flow distributions, and congestion patterns can support network-level evaluation.
Matching aggregate traffic patterns alone does not identify the underlying behavioral mechanism, since different agent settings can produce similar network outcomes\cite{choi2024addressing}.
Future studies should therefore combine evidence across scales and assess performance on observations or scenarios beyond those used for calibration.
For computationally demanding experiments, surrogate models could help explore relationships between agent settings and aggregate outcomes, with candidate results checked through the full simulation \cite{angione2022using}.

\subsection{AI Safety and Systemic Consequences}
The shared transportation environment also provides a setting for studying whether adverse agent behavior produces consequences beyond the agents directly involved \cite{hammond2025multi}.
Future experiments could vary the objectives of agents in different transportation roles, information quality, or the prevalence of deliberately disruptive agents and examine effects on congestion, payments, and the distribution of travel burdens.
These experiments should distinguish infeasible actions blocked by validation from admissible actions that produce harmful collective outcomes.
Trajectory deviations alone do not establish harmful intent or a safety failure; evaluation requires explicit criteria and comparisons accounting for simulation and model variability.
Such future experiments could assess action limits, policy oversight, and other safeguards.

\subsection{Experimental Evidence and Reproducibility}
Each reported configuration has one run, successive days share persistent states, and the MNL coefficients are uncalibrated.
Candidate-based and native MATSim choices also differ in their available action spaces.
The results therefore characterize these configurations without establishing general comparative effectiveness, causal effects, or behavioral realism.
Independent runs with documented simulation and model-generation settings are needed to assess variability.
Comparisons of particular mechanisms can branch from a common pre-intervention state, provided that state restoration is implemented and verified.
Historical reproduction requirements are specified in Appendix~\ref{app:archive-scope}.

\subsection{Extension Beyond the Current Implementation}
The current route--departure and link-charge setting is an initial test case.
Further experiments should vary candidate sets, preference assumptions, and scenarios.
Future extensions include mode, destination, and activity choices, as well as signal control, access restrictions, and transport service provision.
Each extension requires specifying role-specific observations, admissible actions, decision timing, and persistent states, together with backend support for executing the actions and returning relevant feedback.
New roles and shared-resource representations may also be needed.
Integration with other urban activities would require continuity of agent identities and relevant states across those activities.

\section{Conclusion}
\label{sec:conclusion}

AgenticMobility is an open-source experimental platform for studying multi-AI-agent mobility dynamics.
It brings a large population of mobility AI agents together with other AI agents in complementary transportation roles to examine how their behavior, adaptation, and interactions shape collective mobility outcomes.
Persistent identities and states support repeated experience, with traceable interfaces linking decisions, execution, and feedback.
In the current MATSim-based implementation, LLM agents act as travelers choosing routes and departure times and as a manager periodically adjusting link charges; full-day background mobility remains active.

Five configurations completed 100 simulated days each.
Audits of the combined run confirmed action application and personal-memory continuity.
Aggregate trajectories show route-choice stability alongside substantial departure adjustments and differences between controlled-trip and population-wide outcomes.
These results demonstrate the platform's operation and motivate further study; they do not isolate causal effects or establish behavioral realism.

Future experiments can vary information access, retained experience, and management feedback, while extending action spaces and execution interfaces to study richer behaviors and transportation roles.

\section*{Declaration of Generative AI Use}

During the preparation of this manuscript, the authors used ChatGPT (OpenAI) to assist with language editing, improving clarity and readability, and refining the organization and presentation of the manuscript. The research ideas, conceptual framework, literature selection and interpretation, analysis, and conclusions were developed by the authors. All AI-assisted content was reviewed and revised by the authors, who take full responsibility for the content of the manuscript.

\section*{Acknowledgment}
This work is supported by the Natural Science Foundation of China (72671151, 72274024).

% Bibliography
\clearpage
\bibliographystyle{unsrtnat}
\bibliography{references}

\clearpage
\appendix
\section*{APPENDIX}
\addcontentsline{toc}{section}{Appendix}
\section{Detailed Agent and Execution Specifications}
\label{app:specifications}

\subsection{Implemented Mobility Protocol}
\label{app:concrete-protocol}

This subsection specifies the detailed protocol used in the reported demonstration.
Daily rounds, fixed target trips, route--departure choices, and link charges are choices of this implementation.
The general interfaces are defined in Section~\ref{sec:system_framework}.

\subsubsection{Roles and Information}
\label{sec:roles-state-implementation}

The mobility environment is a directed network $G=(V,E)$, where $V$ contains nodes and $E$ contains traversable directed links.
Section~\ref{sec:implemented-setting} describes role-specific decisions and information access.
Agent identities, recent experiences, summary memories, and active policies persist across days; vehicle movements are simulated anew.
A controlled trip's origin, destination, mode, and candidate sets remain fixed.
Travelers do not receive network-wide realized conditions or realized travel times for unchosen routes, and the manager does not receive person-level payments.
Appendix~\ref{app:information-access} specifies detailed input construction and initialization.

\subsubsection{Traveler and Manager Actions}
\label{sec:actions-implementation}

For target trip $i$ on log day $d$, the selected action is
\begin{equation}
 a_{i,d}=(r_{i,d},\tau_{i,d})\in\mathcal{R}_i\times\mathcal{T}_i,
 \label{eq:joint-action}
\end{equation}
where $\mathcal{R}_i$ and $\mathcal{T}_i$ are fixed route and departure-time sets.
The time grid is anchored to the original plan rather than shifted after each choice.
Times $\tau_{i,d}$ are stored in seconds.
A selected departure can differ from realized departure $\hat\tau_{i,d}$ because of preceding activities and trips.
The generated reason is retained separately from the executable action.

The manager submits a complete policy assigning finite, nonnegative charges to eligible directed links.
Unlisted links have zero charge.
A valid policy replaces its predecessor; an empty policy removes all charges.
Between accepted updates, the active policy persists.
Charges remain constant within each day and cover the full execution interval.

Let $p_d(e)$ be the charge for an eligible entry to link $e$.
Actual payment is
\begin{equation}
 C_{i,d}=\sum_{n=1}^{N_{i,d}}p_d(e_{i,d,n}),
 \label{eq:realized-payment}
\end{equation}
where $e_{i,d,n}$ is the link associated with the $n$th eligible entry.
Repeated entries incur repeated charges, and payments already incurred remain recorded if a trip fails to complete.
For an otherwise identical completed route with the same entry counts, changing departure time does not change the direct charge under this all-day pricing protocol.

The manager's objective is to reduce full-day delay measured from completed full-link traversals.
This measure excludes waiting before network entry and has different boundaries from complete trip duration.
Completion, trip time, and payments are recorded separately.

\subsubsection{Daily Execution and Feedback}
\label{sec:daily-cycle-implementation}

On scheduled review days, the manager uses public information available through the preceding day to propose a policy.
Travelers then choose target actions under their information permissions.
Accepted choices are applied after any background replanning and before environmental execution.
The environment runs all trips under the active policy and returns personal outcomes and public summaries to the permitted roles.

Traveler decisions and memory updates occur daily.
Manager memory also updates daily, including during the initial toll-free period, while policy decisions occur only at scheduled opportunities.
Updating manager memory does not change the policy.
The archive links observations, generated outputs, accepted actions, execution, feedback, and state updates.

\subsection{Information Access and Execution Details}
\label{app:implementation}

\subsubsection{Information Access and Initialization}
\label{app:information-access}

Announcement visibility is assigned at the person level.
The LLM chooses only the designated target action; other trips follow background mechanisms.
For target decisions, informed inputs include current candidate charges, whereas uninformed inputs encode current costs as unknown rather than zero.
The LLM Traveler configuration explicitly communicates that charging is disabled.
In the combined configuration, the two information conditions remain distinct even while the active policy is empty.

Initial factual histories and generated summaries are empty.
Synthetic profiles express assumed preferences and are not observations of earlier behavior.
Subsequent traveler inputs combine fixed conditions, up to three preceding days of personal outcomes, the current summary, and permitted announcements.

Manager policy inputs contain public network topology, capacity information, aggregate outcomes, policy history, and manager memory.
The compact input retains the latest two executed policy-version effects and pools older observations.
Recent and preceding summaries cover up to seven observed days.
Detailed link statistics prioritize high-delay or failed links and currently or recently charged links.
An omitted detailed row does not imply that its link is uncongested.
The complete research archive is separate from the input supplied to the manager.

\subsubsection{A Linked Manager Record}
\label{app:manager-trace}

Table~\ref{tab:manager-trace} follows the introduction of charging on Day~50 in the combined run.
The summary written after Day~49 matches the memory in the Day~50 policy input.
That input contains observations through Day~49 and the active zero-charge policy.
The accepted output introduces the two link charges, and the application record identifies them as effective on Day~50.

After execution, the public feedback records 391,340 completed full-link traversals and 33,644 completed car legs, with no aborted or unfinished car legs.
The Day~50 memory update produces a 555-word replacement summary.
Its opening records the start of the priced episode, and its text discusses the first day's outcomes as preliminary observations.
These statements are generated interpretations; the structured feedback remains the measurement record.
The summary matches the previous-memory field in the Day~51 memory-update input.

\begin{table}[!tbp]
\centering\small
\caption{Linked manager records in the combined run. Days use zero-based
log indices.
A memory update does not itself change the active policy.}
\label{tab:manager-trace}
\linespread{1.2}\selectfont
\begin{tabularx}{\linewidth}{@{}>{\raggedright\arraybackslash}p{0.25\linewidth}
>{\raggedright\arraybackslash}X@{}}
\toprule
Stage & Recorded connection \\
\midrule
Before Day~50 execution & Public observations through Day~49; empty charging
policy; summary written after Day~49 \\
Day~50 policy decision & Accepted \texttt{test} action: 1.5 units per entry
on \texttt{43\_3} and \texttt{43\_4} over $[0,108000)$ seconds \\
Day~50 policy application & Active policy identifier
\texttt{pricing-v2-it50} \\
After Day~50 execution & Public feedback enters a complete replacement
summary; the memory update leaves the policy unchanged \\
Day~51 memory input & Contains the summary written after Day~50 \\
Day~80 policy decision & Accepted \texttt{hold} action; policy identifier
changes to \texttt{pricing-v3-it80}, but the price fingerprint is unchanged \\
\bottomrule
\end{tabularx}
\end{table}

The corresponding policy inputs, decisions, application records, and daily memory records are retained in the run archive.
The sequence establishes which feedback and state were available at each step.
Policy retention alone does not establish an improvement or identify the cause of a decision.

\subsubsection{Output Validation and Failure Handling}
\label{app:output-validation}

Traveler choices are checked for the required JSON structure, admissible route and departure values, and text limits.
Memory outputs are checked against their structure and length requirements.
The configured limit is three output attempts: the initial response and up to two retries with validation feedback.
Truncated responses are rejected without reconstructing missing fields or inventing actions.

Persistent traveler-choice or personal-memory failure terminates the experiment.
For manager policy generation, exhausting the allowed invalid-output attempts leaves the current policy in place and records a skipped update.
This fallback concerns invalid generated outputs: model-call exceptions, request-budget failures, or outer execution stalls can still abort a run.
Transport retries and output-validation attempts are separate.

\begin{figure}[!tbp]
\centering
\includegraphics[width=\linewidth,height=0.46\textheight,keepaspectratio]
{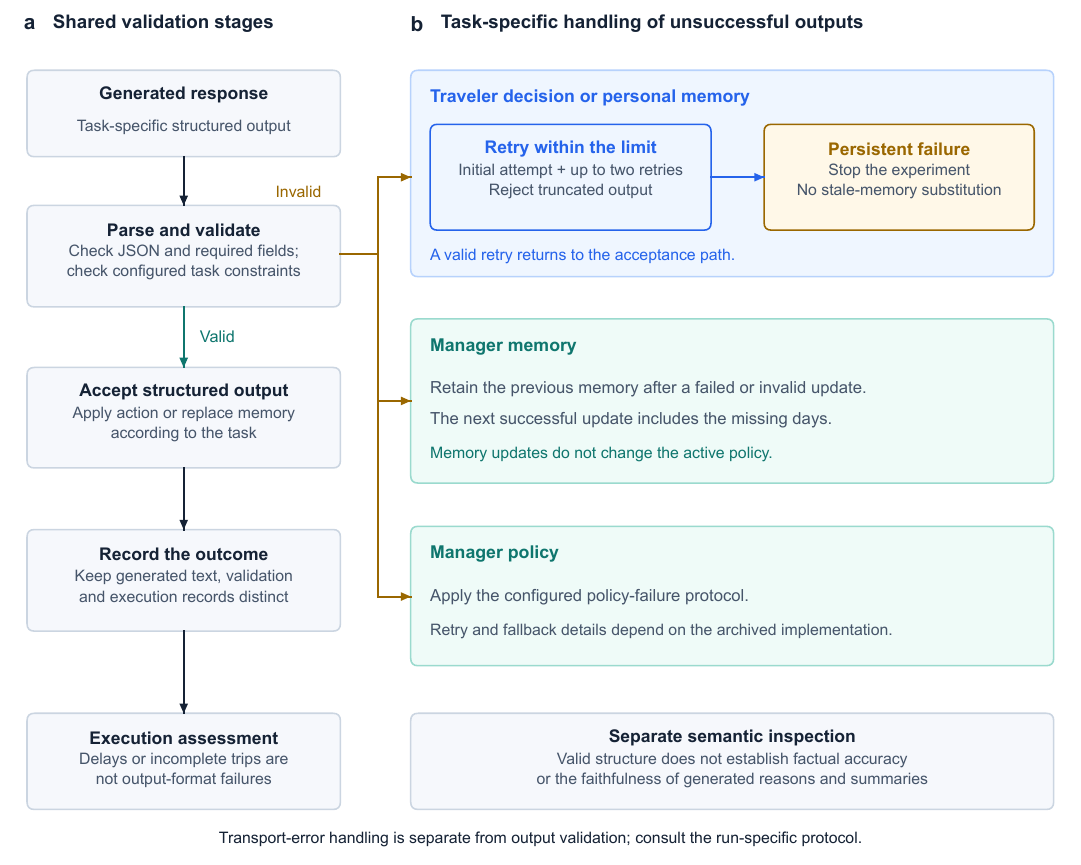}
\caption{Validation and task-specific failure handling. Persistent traveler
task failure stops the run.
Manager policy output failure retains the active policy, and manager-memory failure retains the previous summary.
Runtime and transport failures are handled separately, as described in the text.}
\label{fig:validation_failure}
\end{figure}

Failed, invalid, or timed-out manager-memory updates retain the previous summary and record the fallback.
A later update receives observations since the last successful update.
Cancelled workers cannot commit late memory results.
Figure~\ref{fig:validation_failure} summarizes the paths.
All three policy decisions in the combined demonstration were accepted; the policy fallback was therefore not exercised by those decisions.

Semantic review was disabled in the reported runs.
Structural validity does not certify the factual accuracy of generated memories or the faithfulness of explanations.
Late arrival and noncompletion are environmental outcomes, not automatically output-format failures.

\subsubsection{Plan Execution and Billing}
\label{app:execution-billing}

Accepted choices are written into target plans after any background replanning and before environmental execution.
Original, selected, and realized departure times remain distinct records.
Earlier activities and trips can delay the selected departure.

Charges apply to eligible car-link entries over $[0,108000)$ seconds.
Repeated traversals are charged again, while duplicate notifications of the same entry are ignored.
Candidate cost estimates and realized payments are recorded separately.
Link-delay statistics use completed full-link traversals and have different event boundaries from trip duration and payments.

\subsection{Decision Models and Prompt Specification}
\label{app:decision-models}

\subsubsection{Constrained MNL Baseline}

The joint baseline is identified as \texttt{morning-joint-mnl-v1}.
For candidate action $a=(r,\tau)$, deterministic utility is
\begin{equation}
 V_{i,d}(a)=-0.3\widehat T_{i,d}(a)-0.3\widehat C_{i,d}(a)
 -0.9\widehat L_{i,d}(a)-0.05\lvert\Delta\tau_i(a)\rvert,
 \label{eq:appendix-mnl}
\end{equation}
where $\widehat T$ is expected travel time, $\widehat C$ is expected cost, and $\widehat L$ is predicted lateness.
Time-related terms are in minutes and cost is in scenario units.
With departure times stored in seconds, $\Delta\tau_i(a)=(\tau-\tau_i^0)/60$, where $\tau_i^0$ is the original departure.
Choice probabilities are
\begin{equation}
 P_{i,d}(a)=\frac{\exp\{V_{i,d}(a)\}}
 {\sum_{b\in\mathcal{R}_i\times\mathcal{T}_i}\exp\{V_{i,d}(b)\}}.
\end{equation}
The softmax scale is one.
The homogeneous coefficients are engineering assumptions and are not calibrated to observed behavior.
This baseline supplies a transparent candidate-based reference; its performance does not establish the relative behavioral validity of MNL and LLM decision-making.

When completed personal observations exist within the preceding three-day window for the same route and selected-departure bin, expected time combines static free-flow time and mean observed car time with equal weights.
Otherwise, it uses free-flow time.
Departure bins are 900 seconds wide.
The reported MNL configuration has no tolls and uses neither generated memory summaries nor network-wide traffic observations.

\subsubsection{LLM Settings and Prompt Entry Points}

Model identifiers and generation settings are reported in Table~\ref{tab:llm-settings}; the experience-retention mechanism is described in Section~\ref{sec:agent-implementation}.

The archived template identifiers are:
\begin{itemize}
\item Traveler choice: \texttt{morning-information-v9-continuous-memory}.
\item Personal memory: \texttt{morning-reflection-v9-continuous-memory}.
\item Manager policy: \texttt{morning-pricing-v10-daily-memory}.
\item Manager memory: \texttt{morning-manager-memory-v1-daily}.
\end{itemize}

The implementation stores the traveler-choice and personal-memory prompt definitions in \path{MorningPromptRenderer.java} and \path{MorningMemoryUpdater.java}.
The basic manager policy prompt is \texttt{BASIC\_SYSTEM} in \path{MorningManager.java}, and the manager-memory prompt is in \path{MorningManagerMemory.java}.
These files are located under \path{src/main/java/org/matsim/agentic/} in the repository.
Inputs use declared table columns, numeric times, and explicit unknown values.

Historical attempt records retain the system and user messages, response, schema, and validation information.
They are stored in \path{attempts.jsonl.gz}, \path{memory-attempts.jsonl.gz}, \path{manager-attempts.jsonl.gz}, and \path{manager-memory-attempts.jsonl.gz}.
The source entry points describe the implementation; exact historical inputs must be taken from the corresponding run records.
Template identifiers alone do not replace these inputs.
Appendix~\ref{app:archive-scope} distinguishes the historical archive from the released documentation and excerpts.

\section{Supplementary Experimental Settings and Results}
\label{app:experiments}

\subsection{Demonstration Conditions}
\label{app:demonstration}
\label{sec:experimental-design}
\label{sec:scenario-conditions}

The five configurations in Table~\ref{tab:conditions} use simulation seed 268 and cover log Days~0--99, equivalent to simulation days 1--100.
Population sampling and information assignment use seed 6021.
The fixed target cohort spans 2,877 origin--destination pairs.
The archive labels \texttt{Native}, \texttt{Traveler}, \texttt{Manager}, and \texttt{Joint} denote MATSim, LLM Traveler, MATSim + Manager, and LLM Traveler + Manager.

Manager policy reviews occur on log Days~20, 50, and 80.
MATSim + Manager begins positive charging on Day~20.
The combined configuration retains zero charges at its first review, introduces charges of 1.5 scenario units per entry on links \texttt{43\_3} and \texttt{43\_4} on Day~50, and retains those charges on Day~80.
Its charging period therefore covers Days~50--99.

The combined configuration assigns 5,897 travelers to current announcements and 5,896 to no current announcements.
Both groups receive personal feedback and share the same environment.
Candidate-based configurations use the same fixed alternatives; MATSim agents are not restricted to those sets.
The MNL reference mechanism uses uncalibrated coefficients.
These conditions exercise different decision mechanisms and information settings.
They do not provide matched comparisons isolating the effects of LLM reasoning, summary memory, announcements, or management intervention.

\subsection{Scenario, Candidate Actions, and Configuration}
\label{app:configuration}

\subsubsection{Population and Target Selection}

The input scenario contains 84,110 people.
A fixed 20\% sample selected using seed 6021 yields 16,822 simulated people.
Network flow and storage capacity factors are 0.2 and $0.2^{0.75}\approx0.2991$, respectively \cite{mersiniScalingMATSimTraffic2026,ben-dorPopulationDownscalingMultiagent2021}.

Target trips are selected from original plans using destination-activity start times in $[0,32400)$ seconds, corresponding to starts before 09:00.
Selection covers all activity purposes and requires one car leg per target trip.
The cohort contains 11,793 trips by 11,793 people, with membership fixed throughout the run.
The remaining 5,029 people and non-target trips of controlled people retain background decision mechanisms.
Work-arrival evaluation covers 11,332 targets with work deadlines.
The 09:00 selection threshold does not replace each trip's own deadline.

\subsubsection{Fixed Candidate Sets}

MNL and LLM travelers use the same precomputed BFS-LE route candidate file \cite{rieser2013route}.
The configuration permits up to four routes per OD pair, with maximum pairwise overlap 0.85.
For free-flow time, an alternative passes the detour check if its ratio to the reference is at most 2.5 or its excess time is at most 300 seconds.
The corresponding distance limits are 2.5 and 2,000 meters.
Both the time and distance checks must pass.

Departure candidates span $-5400$ to $+1800$ seconds relative to the original planned departure at 900-second intervals.
Options are clipped against activity and simulation boundaries, and the original departure remains available.
The arrival deadline is not a hard feasibility filter.
Thus, action admissibility concerns the configured candidate space, rather than a guarantee of punctual arrival or minimal schedule displacement.
Of the 11,793 targets, 2,780 have one route and 9,013 have multiple routes.
Single-route targets cannot switch routes within the fixed candidate set.

The principal settings are reported in Table~\ref{tab:configuration}.
The effective innovation cutoff is identified from the run logs and final configuration.
It affects background replanning and MATSim-agent target decisions.
LLM and MNL target choices continue under their respective decision mechanisms.

\subsection{Supplementary Numerical Results}
\label{app:supplementary}

Tables~\ref{tab:validation-outcomes} and~\ref{tab:supp-switching} summarize one run per configuration at seed 268.
They provide numerical reference records for Section~\ref{sec:illustrative-use}.
Display precision does not quantify uncertainty or establish general comparative effectiveness.

\begin{table}[!tbp]
\centering\small
\caption{Mean outcomes over the 100-day window. Daily car time is normalized
by 16,822 people; target time covers 11,793 trips; work lateness covers 11,332 work targets.}
\label{tab:validation-outcomes}
\linespread{1.2}\selectfont
\begin{tabularx}{\linewidth}{@{}>{\raggedright\arraybackslash}Xrrr@{}}
\toprule
Configuration & \shortstack{Daily car time\\(min/person)}
& \shortstack{Target time\\(min/trip)} & \shortstack{Work lateness\\(\%)} \\
\midrule
MATSim & 44.34 & 29.32 & 19.97 \\
MNL & 61.35 & 53.19 & 31.25 \\
LLM Traveler & 35.52 & 17.40 & 5.61 \\
MATSim + Manager & 45.16 & 29.64 & 20.42 \\
LLM Traveler + Manager & 40.71 & 16.25 & 3.73 \\
\bottomrule
\end{tabularx}
\end{table}

\begin{table}[!tbp]
\centering\small
\caption{Mean choice-change rates over 99 transitions. Overall rates use
11,793 targets; multi-route rates use 9,013 targets with multiple fixed candidates.
All values are percentages.
A dash denotes a candidate-space comparison that is not applicable.}
\label{tab:supp-switching}
\linespread{1.2}\selectfont
\begin{tabularx}{\linewidth}{@{}>{\raggedright\arraybackslash}Xrrr@{}}
\toprule
Configuration & \shortstack{Overall\\route changes}
& \shortstack{Multi-route\\changes} & \shortstack{Departure\\changes} \\
\midrule
MATSim & 5.26 & --- & 20.05 \\
MNL & 48.55 & 63.52 & 85.24 \\
LLM Traveler & 3.08 & 4.03 & 29.45 \\
MATSim + Manager & 5.43 & --- & 20.28 \\
LLM Traveler + Manager & 2.19 & 2.86 & 28.04 \\
\bottomrule
\end{tabularx}
\end{table}

A supplementary MATSim and MATSim + Manager comparison at seed 9261 is retained in the archive.
It does not replicate the complete five-configuration experiment and does not change the single-run scope of the tables above.

\section{Audit and Reproducibility Records}
\label{app:records}

\subsection{Detailed Execution Audit and Inference Records}
\label{app:audit-cost}

Table~\ref{tab:execution-audit} summarizes the full-record action and state checks and the sampled information-access inspection.
All 1,179,300 finalized target decisions in LLM Traveler + Manager satisfied the specified action constraints and were linked to applied plans and completed routes.
Output repair affected 2,433 finalized traveler-decision records (0.206\%) and 2,943 personal-memory records (0.250\%).
Personal memory chains remained continuous, and finalized summaries satisfied the configured limit.
The three scheduled manager decisions were valid, and all 100 daily manager-memory updates completed, including one repaired JSON response.

The separate information-access audit inspected 4,000 contexts: 2,000 from LLM Traveler and 2,000 from LLM Traveler + Manager.
No current charging announcements were disclosed to uninformed travelers, and no network-wide traffic information was disclosed in those samples.
This sampled check has a different scope from the full-record action audit.

The combined run contains 1,181,736 traveler-decision calls, 1,182,250 personal-memory calls, three manager-policy calls, and 101 manager-memory calls.
There are 1,179,300 finalized records for each personal task.
Excess calls include retry and repair overhead and are not hallucination counts.
Token totals do not measure monetary cost, energy consumption, or end-to-end runtime; zero LLM calls does not imply zero simulation cost.

\subsection{Run-Specific Reproduction Artifacts}
\label{app:archive-scope}

The historical archive retains configurations, artifact hashes, candidate sets, decision inputs, generated outputs, accepted actions, experiences, policy histories, and memory updates.
The repository documentation supplies aggregate outputs, audit summaries, run metadata, and selected case excerpts.
These materials support different tasks: running the implementation, recalculating summaries, and inspecting recorded interactions.

Historical executables are identified separately for each run.
The authors' review of their differences identified changes in logging or unrelated functionality.
Run identifiers and artifact hashes remain necessary to distinguish those builds from the released source revision.

Regenerating historical trajectories requires the corresponding executable or verified source revision, inputs, prompts, configuration, and model-service settings.
The aggregate outputs and selected excerpts do not contain every historical model response.
The source locations in Appendix~\ref{app:decision-models} identify prompt implementations, while run-specific attempt records preserve the inputs actually used.

% PUBLICATION CHECK: add an existing immutable release tag, commit, or DOI
% in Section 3.4 once assigned. Do not invent it or infer historical
% source-to-executable equivalence from repository HEAD alone.
% The sentence about build differences above relies on the authors'
% confirmation; it is not a new independent binary audit.
\FloatBarrier

\end{document}